\documentclass[
        reprint,
        superscriptaddress,
        aps,prl,
        noeprint,
        altaffilletter
    ]{revtex4-2}

\usepackage{color}
\usepackage{graphicx}
\usepackage{amsmath}
\usepackage{upgreek}
\usepackage{pifont}
\usepackage{siunitx}

\newcommand{\theshallowtrapenergyresolution}{$1.66 {\pm} 0.19$\,eV}

\begin{document}

\title{Tritium Beta Spectrum and Neutrino Mass Limit from Cyclotron Radiation Emission Spectroscopy}

\newcommand{\Washington}{\affiliation{Center for Experimental Nuclear Physics and Astrophysics and Department of Physics, University of Washington, Seattle, WA 98195, USA}}
\newcommand{\Mainz}{\affiliation{Institute for Physics, Johannes Gutenberg University Mainz, 55128 Mainz, Germany}}
\newcommand{\MIT}{\affiliation{Laboratory for Nuclear Science, Massachusetts Institute of Technology, Cambridge, MA 02139, USA}}
\newcommand{\PennState}{\affiliation{Department of Physics, Pennsylvania State University, University Park, PA 16802, USA}}
\newcommand{\PNNL}{\affiliation{Pacific Northwest National Laboratory, Richland, WA 99354, USA}}
\newcommand{\Yale}{\affiliation{Wright Laboratory and Department of Physics, Yale University, New Haven, CT 06520, USA}}
\newcommand{\Livermore}{\affiliation{Lawrence Livermore National Laboratory, Livermore, CA 94550, USA}}
\newcommand{\Case}{\affiliation{Department of Physics, Case Western Reserve University, Cleveland, OH 44106, USA}}
\newcommand{\Indiana}{\affiliation{Center for Exploration of Energy and Matter and Department of Physics, Indiana University, Bloomington, IN, 47405, USA}}
\newcommand{\KIT}{\affiliation{Institute of Astroparticle Physics, Karlsruhe Institute of Technology, 76021 Karlsruhe, Germany}}
\newcommand{\Sorbonne}{\affiliation{Laboratoire de Physique Nucl\'eaire et de Hautes \'Energies, Sorbonne Universit\'e, Universit\'e Paris Cit\'e, CNRS/IN2P3, 75005 Paris, France}}
\newcommand{\CfA}{\affiliation{Center for Astrophysics $\mid$ Harvard $\&$ Smithsonian, Cambridge, MA 02138, USA}}

\author{A.~Ashtari~Esfahani} \altaffiliation{Present Address: Department of Physics, Sharif University of Technology, P.O. Box 11155-9161, Tehran, Iran} \Washington
\author{S.~B\"oser} \Mainz
\author{N.~Buzinsky} \altaffiliation{Present Address: Center for Experimental Nuclear Physics and Astrophysics and Department of Physics, University of Washington, Seattle, WA 98195, USA} \MIT
\author{M.~C.~Carmona-Benitez} \PennState
\author{C.~Claessens} \Washington \Mainz
\author{L.~de~Viveiros} \PennState
\author{P.~J.~Doe} \Washington
\author{M.~Fertl} \Mainz
\author{J.~A.~Formaggio} \MIT
\author{J.~K.~Gaison} \PNNL
\author{L.~Gladstone} \Case
\author{M.~Grando} \PNNL
\author{M.~Guigue} \Sorbonne
\author{J.~Hartse} \Washington
\author{K.~M.~Heeger} \Yale
\author{X.~Huyan} \altaffiliation{Present Address: LeoLabs, Menlo Park, CA 94025, USA} \PNNL
\author{J.~Johnston} \MIT
\author{A.~M.~Jones} \altaffiliation{Present Address: Ozen Engineering, Sunnyvale, CA 94085, USA} \PNNL
\author{K.~Kazkaz} \Livermore
\author{B.~H.~LaRoque} \PNNL
\author{M.~Li} \MIT
\author{A.~Lindman} \Mainz
\author{E.~Machado} \Washington
\author{A.~Marsteller} \Washington
\author{C.~Matth\'e} \Mainz
\author{R.~Mohiuddin} \Case
\author{B.~Monreal} \Case
\author{R.~Mueller} \PennState
\author{J.~A.~Nikkel} \Yale
\author{E.~Novitski} \email{en37@uw.edu} \Washington
\author{N.~S.~Oblath} \PNNL
\author{J.~I.~Pe\~na} \MIT
\author{W.~Pettus} \email{pettus@indiana.edu} \Indiana
\author{R.~Reimann} \Mainz
\author{R.~G.~H.~Robertson} \Washington
\author{D.~Rosa~De~Jes\'us} \PNNL
\author{G.~Rybka} \Washington
\author{L.~Salda\~na} \Yale
\author{M.~Schram} \altaffiliation{Present Address: Thomas Jefferson National Accelerator Facility, Newport News, VA 23606, USA} \PNNL
\author{P.~L.~Slocum} \Yale
\author{J.~Stachurska} \MIT
\author{Y.-H.~Sun} \Case
\author{P.~T.~Surukuchi} \Yale
\author{J.~R.~Tedeschi} \PNNL
\author{A.~B.~Telles} \Yale
\author{F.~Thomas} \Mainz
\author{M.~Thomas} \altaffiliation{Present Address: Booz Allen Hamilton, San Antonio, Texas, 78226, USA} \PNNL
\author{L.~A.~Thorne} \Mainz
\author{T.~Th\"ummler} \KIT
\author{L.~Tvrznikova} \altaffiliation{Present Address: Waymo, Mountain View, CA 94043} \Livermore
\author{W.~Van~De~Pontseele} \MIT
\author{B.~A.~VanDevender} \Washington \PNNL
\author{J.~Weintroub} \CfA
\author{T.~E.~Weiss} \Yale
\author{T.~Wendler} \PennState
\author{A.~Young} \altaffiliation{Present Address: Department of Astrophysics/IMAPP, Radboud University, PO Box 9010, 6500 GL Nijmegen, The Netherlands} \CfA
\author{E.~Zayas} \MIT
\author{A.~Ziegler} \PennState

\collaboration{Project 8 Collaboration}

\date{\today}

\begin{abstract}
The absolute scale of the neutrino mass plays a critical role in physics at every scale, from the particle to the cosmological.
Measurements of the tritium endpoint spectrum have provided the most precise direct limit on the neutrino mass scale.
In this Letter, we present advances by Project 8 to the Cyclotron Radiation Emission Spectroscopy (CRES) technique culminating in the first frequency-based neutrino mass limit.
With only a cm$^3$-scale physical detection volume, a limit of $m_\beta {<} 155$\,eV ($152$\,eV) is extracted from the background-free measurement of the continuous tritium beta spectrum in a Bayesian (frequentist) analysis.
Using $^{83{\rm m}}$Kr calibration data, an improved resolution of \theshallowtrapenergyresolution\ (FWHM) is measured, the detector response model is validated, and the efficiency is characterized over the multi-keV tritium analysis window.
These measurements establish the potential of CRES for a high-sensitivity next-generation direct neutrino mass experiment featuring low background and high resolution.
\end{abstract}

\maketitle


The discovery of neutrino flavor transformation~\cite{Super-Kamiokande:1998kpq,SNO:2002tuh} proves that neutrinos are massive particles, in conflict with the original Standard Model formulation, and establishes that the weak flavor eigenstates ($\nu_e$, $\nu_\upmu$, $\nu_\uptau$) are admixtures of the three neutrino mass eigenstates ($\nu_1$, $\nu_2$, $\nu_3$).
The neutrinos, alone among the fundamental fermions, have masses that remain unmeasured~\cite{ParticleDataGroup:2022pth}.

Neutrino mass is important across nuclear and particle physics, astrophysics, and cosmology.
The origin and magnitude of neutrino mass may hint at new physics~\cite{Petcov:2013poa} such as the neutrino's possible Majorana nature~\cite{Dolinski:2019nrj}, with laboratory searches for neutrinoless double-beta decay~\cite{KamLAND-Zen:2022tow, GERDA:2020xhi, Majorana:2022udl, EXO-200:2019rkq, CUORE:2021mvw} testing this hypothesis.
The tightest, though model-dependent, upper limits on the absolute scale of neutrino mass~\cite{Planck:2018vyg, eBOSS:2020yzd} are derived from measurements of the large-scale structure analyzed within the $\Lambda$CDM cosmological framework~\cite{Lesgourgues:2006nd}.
Neutrino mass has some degeneracy with other parameters, and constraints on it are weakened when allowing additional model extensions~\cite{DiValentino:2015ola}.
Further, emerging cosmological tensions (\emph{e.g.,} $H_0$ determination~\cite{Planck:2018vyg,Riess:2021jrx,Knox:2019rjx}) might point to new physics~\cite{Abdalla:2022yfr}, highlighting the value of independent measurements of $\Lambda$CDM parameters.

A direct and model-independent laboratory constraint on the neutrino mass can be derived from the kinematics of beta decay or electron capture~\cite{Drexlin:2013lha,Formaggio:2021nfz}.
The electron-weighted neutrino mass ($m_{\beta}$) observable is
\begin{equation}
    m_{\beta} = \sqrt{\sum_{i=1}^3 \left| U_{ei} \right|^2 m_i^2},
\end{equation}
where $m_{i=1,2,3}$ are the neutrino mass eigenvalues and $U_{ei}$ are elements of the $3 \times 3$ unitary leptonic mixing matrix~\cite{Pontecorvo:1957cp,Maki:1962mu}.
The effect of neutrino mass manifests in the decay electron spectrum near the endpoint as both a shape distortion and a reduction in maximum electron energy.
Neutrino flavor oscillation measurements, sensitive only to the mass-squared differences, impose an ultimate lower bound of $m_\beta {\geq} 0.009$\,eV/$c^2$ ($m_\beta {\geq} 0.048$\,eV/$c^2$) for the normal (inverted) mass ordering~\cite{ParticleDataGroup:2022pth}.

For over 70 years, tritium beta decay experiments have produced the most sensitive direct $m_\beta$ limits~\cite{Formaggio:2021nfz}.
Most recently, KATRIN has set a limit of $m_\beta {<} 0.8$\,eV/$c^2$ (90\% C.L.)~\cite{KATRIN:2021uub}.
Such experiments using molecular tritium ($^3$H$_2$, or T$_2$) become systematics-limited at $m_\beta {\sim} 0.1$\,eV/$c^2$ due to broadening caused by internal molecular motion~\cite{Saenz:2000dul,Bodine:2015sma}.
An alternative method uses electron capture on $^{163}$Ho~\cite{Gastaldo:2017edk,Alpert:2014lfa}, with a current limit of $m_\beta {<} 150$\,eV/$c^2$~\cite{Velte:2019jvx}; the challenges of complex atomic and solid-state structure, backgrounds, and pileup in this method are being investigated.

The Project 8 Collaboration has developed a new technique, Cyclotron Radiation Emission Spectroscopy (CRES), in pursuit of eventual sensitivity to $m_{\beta}$ down to 0.04\,eV/$c^2$.
CRES uses the cyclotron emission from electrons or positrons to determine their energies~\cite{Monreal:2009za, Asner:2014cwa, Byron:2022wtr}.
The cyclotron frequency ($f_\mathrm{c}$) of electrons in a magnetic field ($B$) is a proxy for their kinetic energy ($E_{\rm{kin}}$):
\begin{equation}
f_\mathrm{c} = \frac{1}{2\pi}\frac{\lvert e \rvert B}{m_e+E_{\rm{kin}}/c^2},
\label{eq:energytofrequency}
\end{equation}
where $e$ is the electron charge, $m_e$ is the mass of the electron, and $c$ is the speed of light in vacuum.
For magnetic field strength of $\mathcal{O}$(T), an electron's faint $\mathcal{O}$(fW) cyclotron radiation can be detected directly at radio frequency (RF).
The high precision possible with a frequency measurement and CRES's inherent relative immunity to background make the technique ideally suited to studying electrons emitted in beta decay~\cite{Esfahani:2017dmu}.
CRES has the statistical advantage of simultaneous acquisition across the energy range of interest and elimination of systematic effects associated with integral spectroscopic methods~\cite{KATRIN:2021fgc, Project8:2022wqh}.
Project 8 first demonstrated this technique on the conversion electrons emitted by gaseous $^{83{\rm m}}$Kr~\cite{Asner:2014cwa}. 
Here we present the first continuous-spectrum measurement using the CRES technique\textemdash on molecular tritium beta decay near the endpoint region\textemdash enabling the first neutrino mass limit using CRES.
We further demonstrate eV-scale resolution and modeling of the detector response using $^{83{\rm m}}$Kr data.
The full details of these analyses are presented in a companion manuscript~\cite{tritiumPRC:2022}.

\begin{figure}[t!]
  \centering
  \includegraphics[width=\columnwidth]{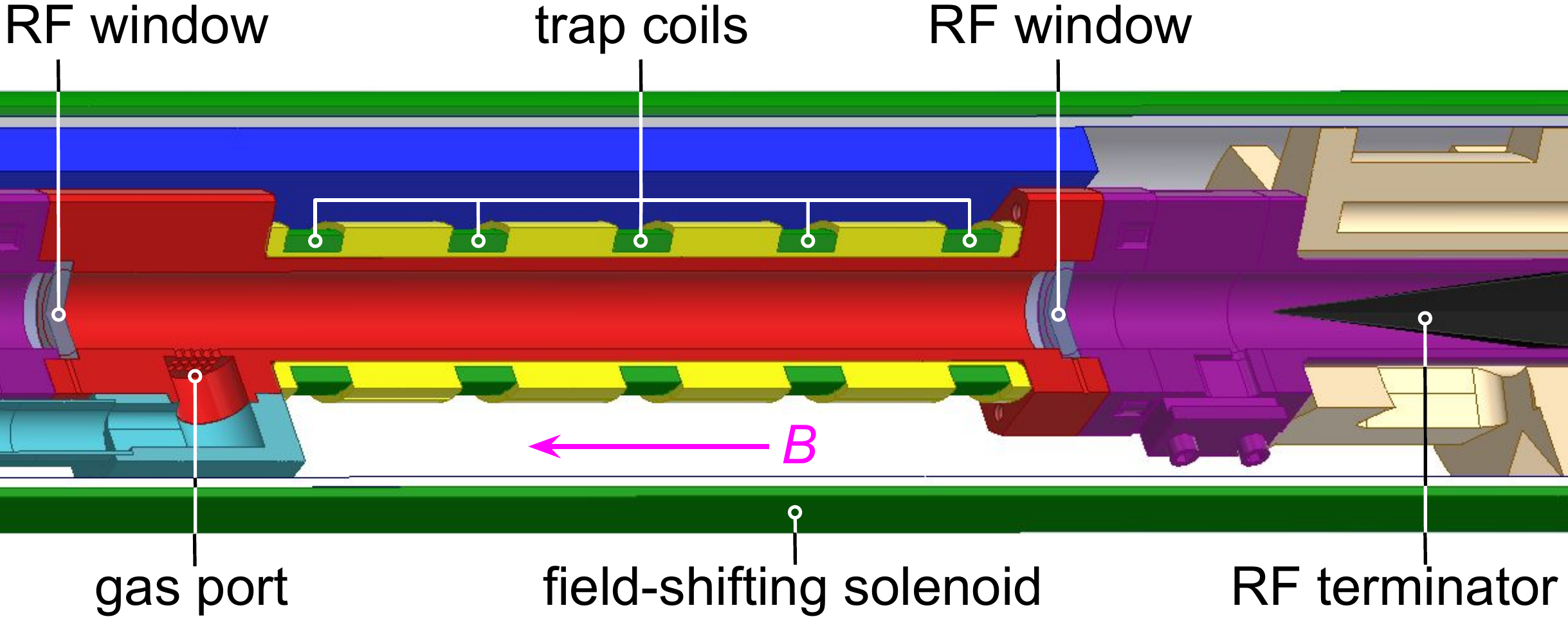}
  \caption{Cutaway of the cryogenic CRES cell, where electrons are produced in radioactive decay and magnetically trapped. Cyclotron radiation travels axially up the waveguide (left in rotated view), toward the amplifiers and readout electronics.
  }
  \label{fig:apparatus}
\end{figure}

At the core of the CRES apparatus is a cryogenic gas cell (the ``CRES cell''), within which electrons are produced in radioactive decay and magnetically trapped while they emit cyclotron radiation (Fig.~\ref{fig:apparatus}).
Radioactive source gas is delivered to this cylindrical cell through an array of sub-wavelength holes and the gas is confined axially by microwave-transparent CaF$_2$ vacuum windows. 
Cooling the cell to 85\,K reduces thermal noise while maintaining sufficient $^{83{\rm m}}$Kr in the gas phase.
The cell is positioned in a superconducting magnet, whose 0.959\,T axial background magnetic field induces cyclotron motion and confines electrons radially.
A ``field-shifting solenoid'' within the magnet bore enables shifting of this background field by up to $\pm$0.3\% for systematic studies.
Five coils wound around the cell provide near-harmonic magnetic trap potentials to confine electrons axially.
Data were taken in composite traps: a ``shallow trap'' of two coils with depth ${\sim}0.08$\,mT to demonstrate high-resolution CRES and a ``deep trap'' of four coils with depth ${\sim}1.4$\,mT to increase effective volume for the tritium endpoint measurement.

Source gases\textemdash molecular tritium and $^{83{\rm m}}$Kr\textemdash are delivered individually from a custom gas manifold.
Tritium is stored in a non-evaporable getter, with its operating partial pressure stabilized at $10^{-6}$\,mbar using a feedback-loop-controlled heating current to optimally balance the event rate with the rate of unwanted electron-gas collisions.
In a separate calibration mode, $^{83{\rm m}}$Kr emanates from $^{83}$Rb adsorbed in zeolite~\cite{Venos:2005vn}, and the rate of electron collisions is tuned to match that in tritium data by the controlled release of H$_2$ from a separate getter.
The gas composition is monitored using quadrupole mass analyzers.
To remove unwanted $^3$He from the decay of tritium adsorbed on gas system walls, in later data sets the gas from the active volume is continuously pumped through a controllable leak valve and sequestered.

The CRES cell is a circular waveguide segment with an inner diameter of 10.06\,mm.
The 26\,GHz cyclotron radiation couples to the TE$_{11}$ propagating mode.
Only the upward-propagating radiation is detected; it is transmitted via waveguide to a series pair of low-noise cryogenic amplifiers held at 30\,K.
To avoid unwanted reflections, the downward-propagating radiation is absorbed below the cell in a conical graphite-epoxy terminator, thermal noise from which is the dominant contributor to the 132$\pm$7\,K system noise temperature.
After amplification, the signal is mixed down in frequency and filtered before being digitized by a ROACH2 data acquisition (DAQ) system~\cite{Hickish2016} sampling at 3.2\,GS/s.
The onboard FPGA performs Fourier transforms and digital downconversion to 200\,MS/s in three independently-tunable DAQ frequency windows.
The data are streamed to a compute node that applies trigger logic, writing periods of time-series data to disk based on high observed power in frequency space.

\begin{figure}[b!]
  \centering
  \includegraphics[width=\columnwidth]{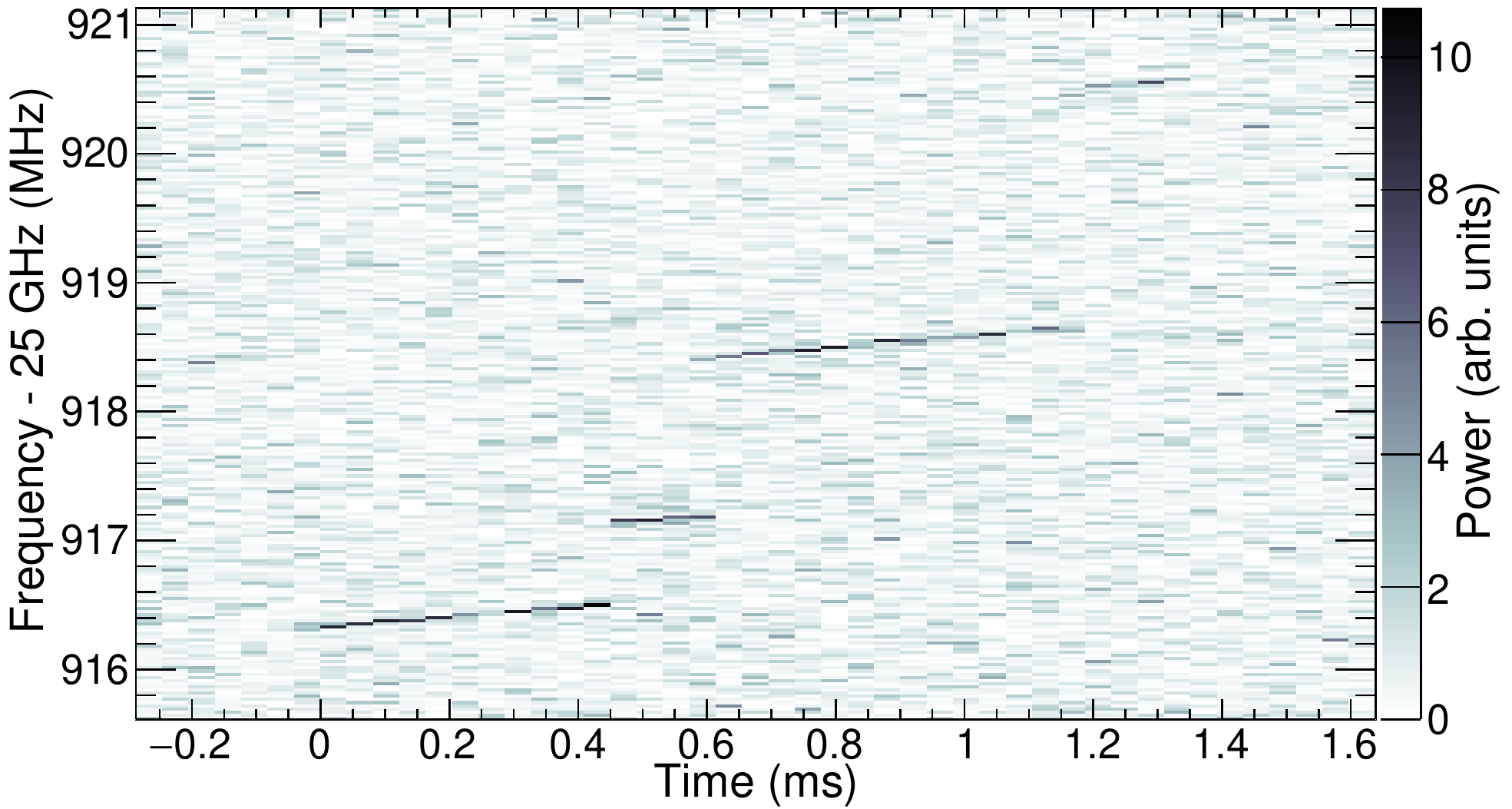}
  \caption{
    Spectrogram of the first CRES event detected from a tritium beta decay electron.
    Raw time-series data are Fourier-transformed in time bins of 40.96\,$\upmu$s, yielding 24.41\,kHz frequency bins.
  }
  \label{fig:spectrogram}
\end{figure}

The Fourier-transformed data form a two-dimensional spectrogram of power as a function of frequency and time (Fig.~\ref{fig:spectrogram}).
A tunable point-clustering algorithm is used to identify bins with a high signal-to-noise ratio (SNR) that belong to electron signals and to group these into contiguous ``tracks.''
Tracks are always ``chirped''\textemdash positive-sloped in frequency\textemdash due to radiated cyclotron power (Eq.~\ref{eq:energytofrequency}).
Inelastic collisions between electrons and gas molecules cause energy loss and therefore sudden jumps in cyclotron frequency, so tracks in close time sequence are designated as being a single electron ``event.''
An electron's energy at the time of decay is extracted from the initial frequency of the earliest track in an event.
The on-axis detector geometry introduces a Doppler shift, shunting power from the main carrier into sidebands and thereby limiting the SNR and effective volume.
All tracks in the spectrogram in Fig.~\ref{fig:spectrogram} are the main carrier signal; sideband tracks due to modulation are present at frequencies above and below the main carrier but are too low-power to be detected in this apparatus~\cite{Esfahani:2019mpr}.

CRES is inherently an extremely low-background technique, with RF noise fluctuations as the dominant background.
By precisely characterizing the RF background, true events can be sensitively distinguished from noise.
In the event selection, the decision to keep or remove an event is based on the number of tracks it contains and properties of its first track: the number of high-power bins it contains (roughly corresponding to its duration) and its average SNR.
The parameters for this cut were chosen before tritium data acquisition at a level expected to allow less than one RF-noise-induced background event in the tritium data set with 90\% probability.


Analysis is performed on binned cyclotron frequency ($f_\mathrm{c}$) data~\cite{tritiumPRC:2022}.
The predicted spectral shape ($\mathcal{S}$) as detected in a CRES apparatus is
\begin{equation}
    \mathcal{S}= \epsilon \left(\mathcal{Y}* \sum_{j=0}\mathcal{A}_j\left(\mathcal{I}*\mathcal{L}^{*j}\right)\right),
    \label{eq:FullModel}
\end{equation}
where all variables can be expressed as functions of $f_\mathrm{c}$, and consequently $E_{\mathrm{kin}}$ and $B$ (Eq.~\ref{eq:energytofrequency}).
Convolution is denoted by $*$, with superscript $*j$ representing self-convolution $j$ times.
The detection efficiency ($\epsilon$) uniquely has explicit physical dependence on both $f_\mathrm{c}$ and $E_{\mathrm{kin}}$.
The true underlying spectrum ($\mathcal{Y}$) includes all source effects.
The summation term characterizes the broadening elements of the detector response, primarily from scattering and the inhomogeneity of the magnetic trapping field.
Scattering before an electron’s detection gives rise to a low-energy tail populated by events with undetected true first tracks.
Scatter peak amplitudes ($\mathcal{A}_j$) are the probabilities that an electron is first detected after $j$ scatters.  They are determined by a phenomenological model.
The electron's energy-loss distribution after $j$ scatters ($\mathcal{L}^{*j}$) depends on the differential cross sections, the fractions $\gamma_i$ of each gas species $i$, and the loss to cyclotron radiation. 
The intrinsic instrumental resolution ($\mathcal{I}$) arises from variation in the average magnetic fields sampled by electrons with different kinematic parameters~\cite{Esfahani:2019mpr}.
It is modeled using simulations of monoenergetic electrons in the trap's magnetic field profile~\cite{AshtariEsfahani:2019mwv}.

\begin{figure}[t]
  \centering
  \includegraphics[width=1.0\columnwidth]{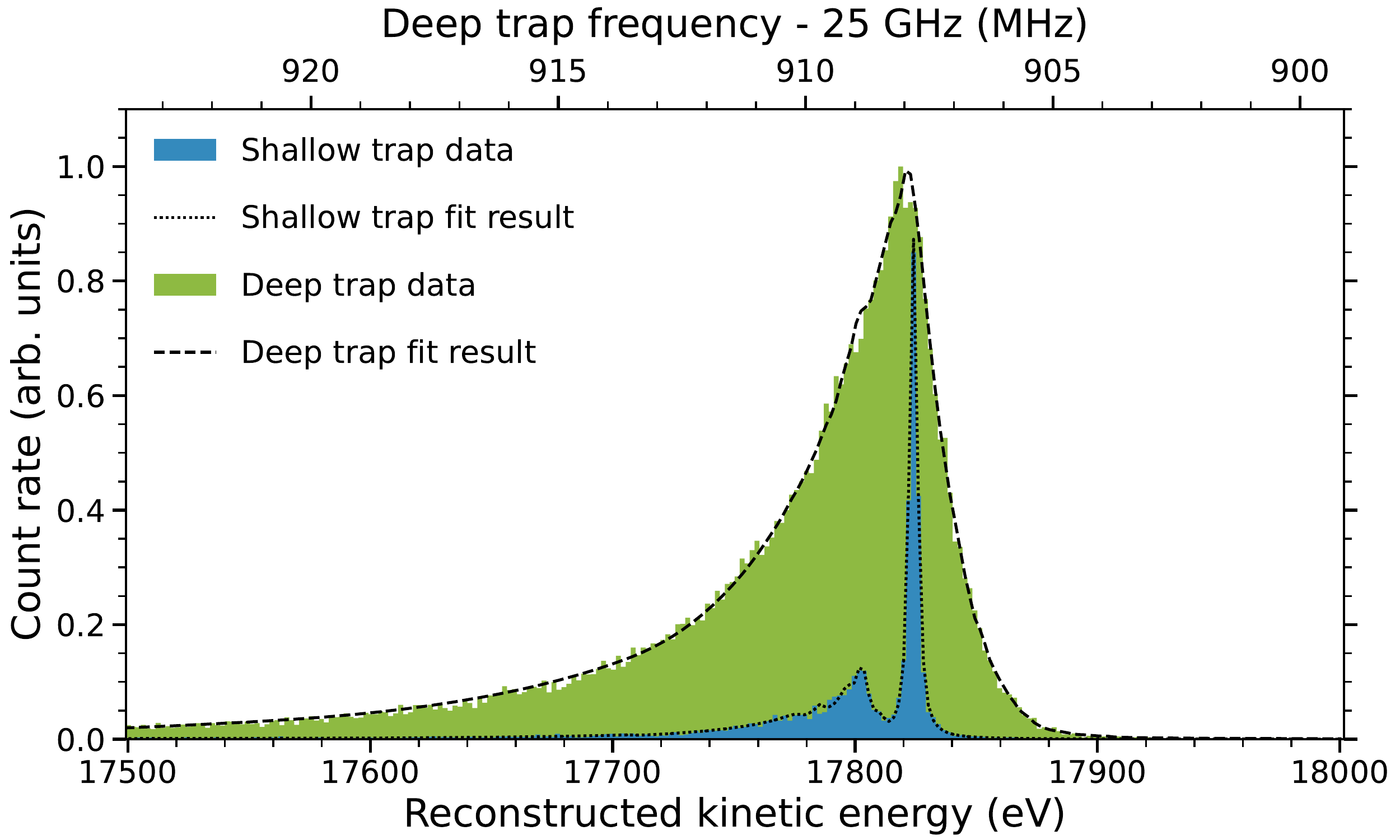}
  \caption{
    Data and fits of the 17.8\,keV $^{83{\rm m}}$Kr conversion electron K-line, as measured in the shallow (high-resolution) and the deep (high-statistics) electron trapping configurations.
    The shallow trap exhibits an instrumental resolution of \theshallowtrapenergyresolution\ (FWHM), while the deep trap provides direct calibration of the tritium data-taking conditions.
  }
  \label{fig:krypton}
\end{figure}

Near-monoenergetic conversion electrons from $^{83{\rm m}}$Kr are a powerful tool for characterizing the detector response near the tritium endpoint.
The underlying spectrum of the $^{83{\rm m}}$Kr K-line at 17.8 keV ($\mathcal{Y}_{\mathrm{Kr}}$) consists of a narrow main peak and secondary low-energy peaks from shakeup and shakeoff~\cite{Robertson:2020dd}.
$^{83{\rm m}}$Kr calibration data (Fig.~\ref{fig:krypton}) are taken in the shallow and deep trap configurations to verify the validity of the detector response model across different trap depths and scattering environments.
Poisson-likelihood $\chi^2$ fits~\cite{Baker:1983tu} are performed on these data, with goodness-of-fit tested using Monte Carlo methods.
The scattering parameters in $\mathcal{A}_j$ and the magnetic field $B$ are free fit parameters, and no background is observed.

In the shallow trap configuration, resolution is optimized by minimizing the magnetic field variation experienced by trapped electrons.
The observed primary peak width is 4.0\,eV (FWHM); the K-line natural linewidth is 2.774\,eV~\cite{Altenmuller:2019ddl}, yielding an instrumental resolution $\mathcal{I}$ of \theshallowtrapenergyresolution.
The low-energy satellite peak consists of overlapping contributions from scattering (61\% of satellite peak counts) and shakeup/shakeoff components.
Across the full shallow-trap $^{83{\rm m}}$Kr spectrum, 69\% of events are detected before scattering.

The event rate is approximately $40$ times higher in the deep trap configuration than in the shallow trap, at the expense of a broader peak width of 54.3\,eV (FWHM).
This deep-trap configuration is also used for data acquisition with tritium, making these $^{83{\rm m}}$Kr extracted parameters inputs for the tritium analysis:
$B {=} 0.9578104(13)$\,T is used directly, while $\mathcal{A}_j$ is corrected for small differences in scattering environment between data sets.

\begin{figure}[t]
  \centering
  \includegraphics[width=1.0\columnwidth]{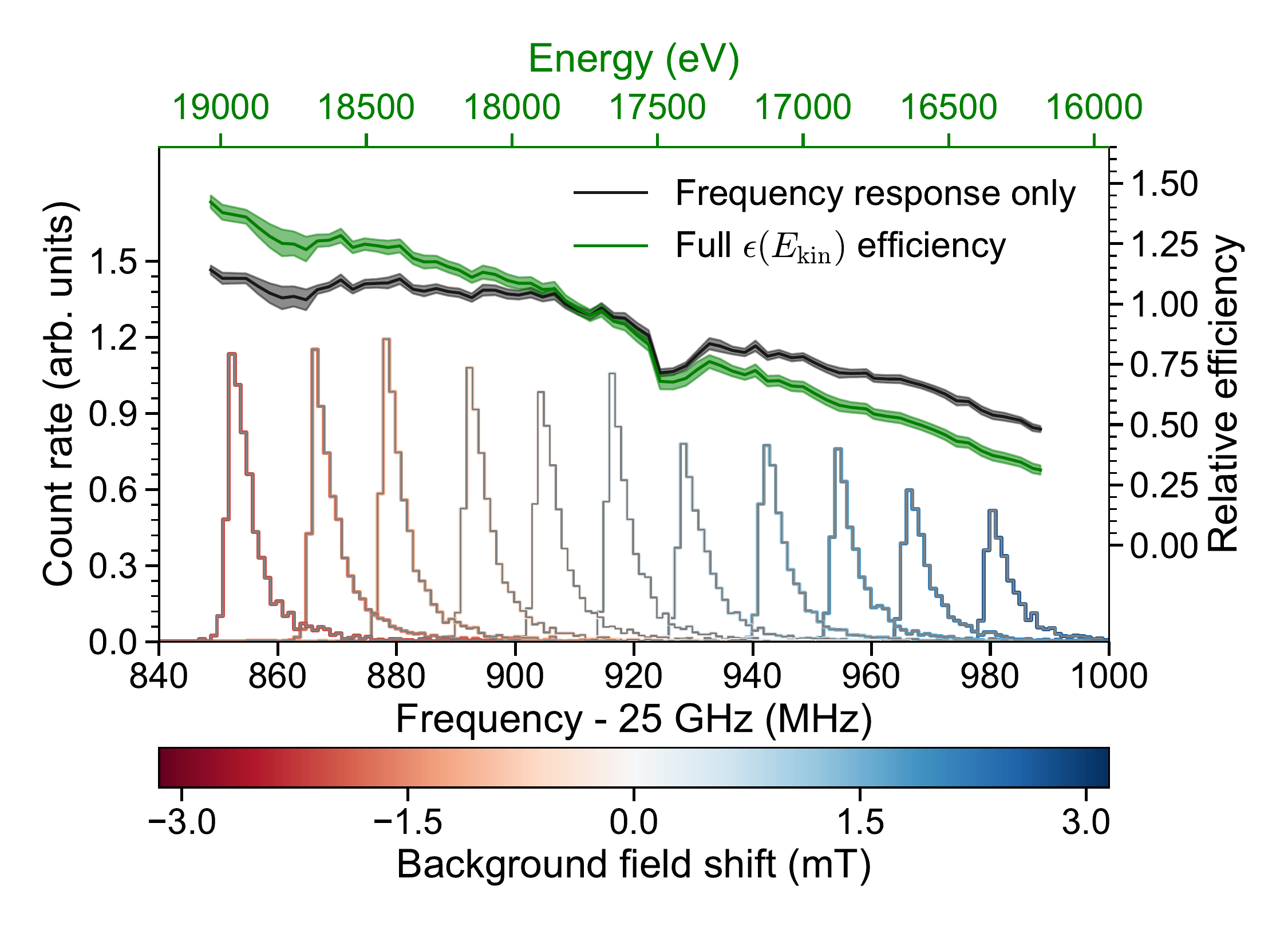}
  \caption{
    The 17.8\,keV $^{83{\rm m}}$Kr conversion electron line recorded in the deep trap with varying magnetic background fields (red to blue). The gray curve shows the efficiency's
    response to frequency variation, extrapolated from single trap data. The green curve is corrected for energy dependence of emitted cyclotron power and shows the relative efficiency predicted for tritium data.
  }
  \label{fig:efficiency}
\end{figure}

The $^{83{\rm m}}$Kr data are also used to measure the frequency variation of efficiency $\epsilon$ and detector response elements $\mathcal{I}$ and $\mathcal{A}_j$.
By varying the background magnetic field in steps of $0.07$\,mT over a range of $\pm$3.2\,mT using the field-shifting solenoid, deep trap $^{83{\rm m}}$Kr 17.8-keV data (Fig.~\ref{fig:efficiency}) are produced at a range of frequencies in the region of interest.
The ``notch'' in detection efficiency at 25.93\,GHz arises from electron interactions with the TM$_{01}$ cavity mode due to reflections from waveguide elements.
Electrons in resonance with this mode lose energy to cyclotron radiation faster than in free space~\cite{Purcell1946, Gabrielse:1985zz}, which increases frequency chirp (track slope) of the CRES signal and reduces the efficiency of the event reconstruction procedure, which has been optimized for the non-enhanced slopes.
In addition to the frequency-dependent effects described above, the efficiency $\epsilon$ for tritium data also includes an analytic term that directly depends on electron energy, in order to account for the radiated power~\cite{Esfahani:2019mpr}.
Detection efficiencies are determined with uncertainties of 2--6\%.
The efficiency $\epsilon$ and the dependence of $\mathcal{A}_j$ on frequency, which are derived from these field-shifted $^{83{\rm m}}$Kr data, are passed as inputs to the tritium data analysis, while variations in $\mathcal{I}$ were found to have negligible impact.

The electron spectrum from tritium beta decay extends out to its endpoint of $E_0{=}18574$\,eV~\cite{Myers:2015lca, Bodine:2015sma, KATRIN:2021uub}.
All three DAQ frequency windows were used simultaneously to record around the endpoint, with the combined analysis window spanning 16.2--19.8\,keV (25.81--25.99\,GHz).
Over the 82-live-day data-taking period, using the high-statistics deep trap configuration, 3770 distinct tritium events were recorded.

The tritium analysis follows Eq.~\ref{eq:FullModel}, where the underlying spectrum $\mathcal{Y}_{\mathrm{tritium}}$~\cite{Kleesiek:2018mel} here is an approximated beta spectrum~\cite{AshtariEsfahani:2021moh} convolved with the final state distribution of the $^3$HeT$^+$ decay product~\cite{Saenz:2000dul}.
A flat background component is included as a free fit parameter.
Approximations are made to the instrumental resolution $\mathcal{I}$ and energy-loss $\mathcal{L}$ to reduce computing time, to account for differences in scattering environment, and to include an explicit parameter $\sigma$ describing instrumental resolution width.
The approximate model produces correct coverages and no biases for ensembles of Monte Carlo data generated with an un-approximated model.

The endpoint and neutrino mass limit are determined using both Bayesian and frequentist analyses validated with Monte Carlo studies.
Separate fits were performed to measure $E_0$ (with $m_\beta$ constrained near 0\,eV) and constrain $m_\beta$ (with $E_0$ floating).
The frequentist analysis best-fit interval for $m_{\beta}^2$ is constructed using the procedure in \cite{Kraus:2004zw}, and is converted to the $m_{\beta}$ limit using the Feldman-Cousins method~\cite{Feldman:1997qc}.

\begin{figure}[t]
  \centering
  \includegraphics[width=1.0\columnwidth]{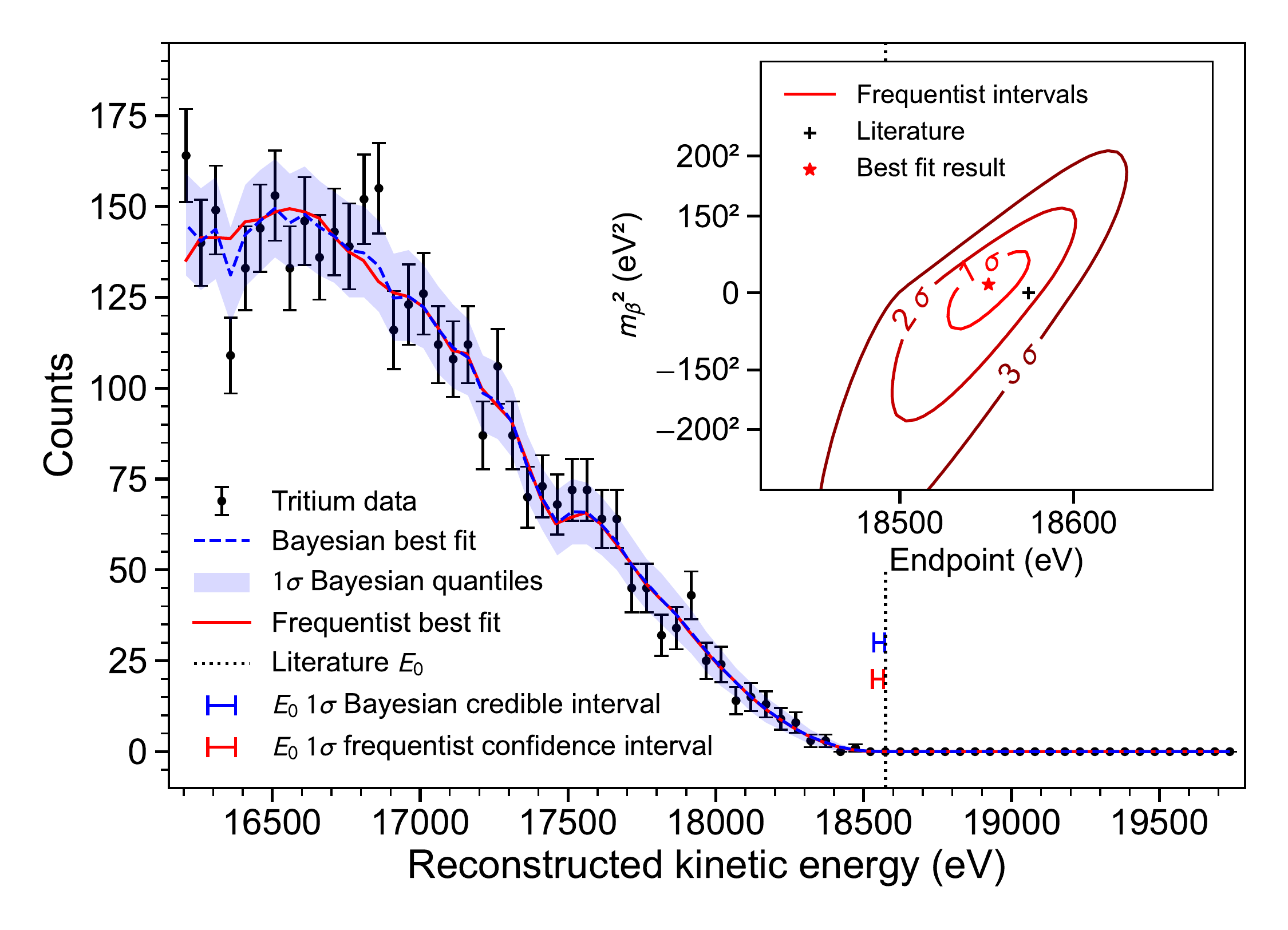}
  \caption{
    Measured tritium endpoint spectrum with Bayesian and frequentist fits. (Inset) Frequentist neutrino mass and endpoint contours.
  }
  \label{fig:tritium}
\end{figure}

Fig.~\ref{fig:tritium} shows the measured tritium spectrum and fits, with results summarized in Tab.~\ref{tab:results}.
The endpoint and mass values are consistent with each other and with literature values.
No counts were detected above the endpoint, setting a stringent upper limit on backgrounds of ${\leq} 3 \times 10^{-10}$\,counts/eV/s.
The endpoint uncertainty contributions are listed in Tab.~\ref{tab:uncertainties}.
Statistical uncertainty dominates the uncertainty budget, with determination of systematic effects also statistics-limited.

\begin{table}[t]
\caption{Extracted tritium endpoint values with 1-$\sigma$ uncertainty and neutrino mass 90\% credible/confidence intervals.  The literature value is $E_0 {=} 18574$\,eV~\cite{Myers:2015lca, Bodine:2015sma, KATRIN:2021uub}.}
\label{tab:results}
\centering
\renewcommand{\arraystretch}{1.5}
\begin{tabular}{p{0.2\columnwidth}>{\centering}p{0.35\columnwidth}>{\centering\arraybackslash}p{0.35\columnwidth}}
\hline\hline
 & Endpoint [eV] & $m_{\beta}$ limit [eV/$c^2$] \\
\hline
Bayesian & $18553^{+18}_{-19}$ & $<$155 \\
Frequentist & $18548^{+19}_{-19}$ & $<$152 \\
\hline\hline
\end{tabular}
\end{table}

\begin{table}[t]
\caption{Contributions to endpoint uncertainty $\sigma(E_0)$ in the frequentist analysis. Individual systematic uncertainty contributions were computed via the method of Asimov sets~\cite{Cowan:2010js}; they therefore do not sum in quadrature to the total systematic uncertainty, which takes into account correlations.}
\label{tab:uncertainties}
\centering
\renewcommand{\arraystretch}{1.1}
\begin{tabular}{p{0.48\columnwidth}>{\centering}p{0.22\columnwidth}>{\centering\arraybackslash}p{0.12\textwidth}}
\hline\hline
Uncertainty & Parameters & $\sigma(E_0)$~[eV] \\
\hline
Magnetic field & $B$ & 4 \\
Magnetic field broadening & $\sigma$ & 4 \\
Scattering & $\gamma_{\mathrm{H}_2}$, $\mathcal{A}_j$ & 6 \\
Efficiency variation & $\epsilon$ & 4 \\
Other freq.~dependence & $\sigma(f_c)$, $\mathcal{A}_j(f_c)$ & 6 \\
\hline
Systematics total & & 9 \\
\hline
Statistical &  & 17 \\
\hline\hline
\end{tabular}
\end{table}


These results highlight the capabilities of the frequency-based CRES technique.
Krypton calibration data demonstrate its inherently high resolution, enabling the full decomposition of the detector response.
Energy- and frequency-dependent effects are measured and controlled to allow analysis across a multi-keV continuous spectrum.
The dominant background is RF noise fluctuations, consistent with expectation, which is characterized and rejected to achieve a zero-background measurement.
These characteristics combine to enable the first tritium endpoint measurement and direct neutrino mass limit with the novel CRES technique.

These measurements demonstrate significant advances for CRES and suggest avenues for improving its sensitivity to $m_\beta$.
The analysis is statistics-limited, motivating pursuit of a large-volume CRES apparatus~\cite{Project8:2022wqh}.
The planned cavity-based detection geometry will benefit from increased signal power due to enhanced spontaneous emission on resonance~\cite{Purcell1946, Gabrielse:1985zz} while also reducing the Doppler shift, thus simplifying event morphology.
Paired with reduced noise, potentially from the use of quantum amplifiers, the SNR and thus detection efficiency can be significantly enhanced.
More sophisticated reconstruction techniques, including matched filtering and/or machine learning~\cite{AshtariEsfahani:2019mxy}, have the potential to further increase reconstruction efficiency and enable the identification of sidebands, providing input for kinematic corrections to improve resolution~\cite{Esfahani:2019mpr}.
Novel calibration with a tunable monoenergetic electron source will be required to further improve detector response characterization, as the CRES resolution has already surpassed the natural linewidth of $^{83{\rm m}}$Kr and atomic shakeup/shakeoff satellites contribute significantly to the $^{83{\rm m}}$Kr lineshape~\cite{Robertson:2020dd}.

Project 8 aims to combine these advances with an atomic tritium source to bypass the molecular final state broadening and uncertainties.
This sets the stage for a next-generation neutrino mass experiment probing the full range of m$_\beta$ allowed by the inverted neutrino mass ordering.

\noindent\textbf{Acknowledgments}

This material is based upon work supported by the following sources: the U.S. Department of Energy Office of Science, Office of Nuclear Physics, under Award No.~DE-SC0020433 to Case Western Reserve University (CWRU), under Award No.~DE-SC0011091 to the Massachusetts Institute of Technology (MIT), under Field Work Proposal Number 73006 at the Pacific Northwest National Laboratory (PNNL), a multiprogram national laboratory operated by Battelle for the U.S. Department of Energy under Contract No.~DE-AC05-76RL01830, under Early Career Award No.~DE-SC0019088 to Pennsylvania State University, under Award No.~DE-FG02-97ER41020 to the University of Washington, and under Award No.~DE-SC0012654 to Yale University; the National Science Foundation under Award No.~PHY-2209530 to Indiana University, and under Award No.~PHY-2110569 to MIT; the Cluster of Excellence “Precision Physics, Fundamental Interactions, and Structure of Matter” (PRISMA+ EXC 2118/1) funded by the German Research Foundation (DFG) within the German Excellence Strategy (Project ID 39083149); the Karlsruhe Institute of Technology (KIT) Center Elementary Particle and Astroparticle Physics (KCETA); Laboratory Directed Research and Development (LDRD) 18-ERD-028 and 20-LW-056 at Lawrence Livermore National Laboratory (LLNL), prepared by LLNL under Contract DE-AC52-07NA27344, LLNL-JRNL-838683; the LDRD Program at PNNL; Indiana University; and Yale University.  Portions of the research were performed using the Core Facility for Advanced Research Computing at CWRU, the Engaging cluster at the MGHPCC facility, Research Computing at PNNL, and the HPC cluster at the Yale Center for Research Computing.  The $^{83}$Rb/$^{83{\rm m}}$Kr isotope used in this research was supplied by the United States Department of Energy Office of Science through the Isotope Program in the Office of Nuclear Physics.

\bibliographystyle{apsrev4-2}
\bibliography{tritium}

\begin{thebibliography}{50}%
\makeatletter
\providecommand \@ifxundefined [1]{%
 \@ifx{#1\undefined}
}%
\providecommand \@ifnum [1]{%
 \ifnum #1\expandafter \@firstoftwo
 \else \expandafter \@secondoftwo
 \fi
}%
\providecommand \@ifx [1]{%
 \ifx #1\expandafter \@firstoftwo
 \else \expandafter \@secondoftwo
 \fi
}%
\providecommand \natexlab [1]{#1}%
\providecommand \enquote  [1]{``#1''}%
\providecommand \bibnamefont  [1]{#1}%
\providecommand \bibfnamefont [1]{#1}%
\providecommand \citenamefont [1]{#1}%
\providecommand \href@noop [0]{\@secondoftwo}%
\providecommand \href [0]{\begingroup \@sanitize@url \@href}%
\providecommand \@href[1]{\@@startlink{#1}\@@href}%
\providecommand \@@href[1]{\endgroup#1\@@endlink}%
\providecommand \@sanitize@url [0]{\catcode `\\12\catcode `\$12\catcode
  `\&12\catcode `\#12\catcode `\^12\catcode `\_12\catcode `\%12\relax}%
\providecommand \@@startlink[1]{}%
\providecommand \@@endlink[0]{}%
\providecommand \url  [0]{\begingroup\@sanitize@url \@url }%
\providecommand \@url [1]{\endgroup\@href {#1}{\urlprefix }}%
\providecommand \urlprefix  [0]{URL }%
\providecommand \Eprint [0]{\href }%
\providecommand \doibase [0]{https://doi.org/}%
\providecommand \selectlanguage [0]{\@gobble}%
\providecommand \bibinfo  [0]{\@secondoftwo}%
\providecommand \bibfield  [0]{\@secondoftwo}%
\providecommand \translation [1]{[#1]}%
\providecommand \BibitemOpen [0]{}%
\providecommand \bibitemStop [0]{}%
\providecommand \bibitemNoStop [0]{.\EOS\space}%
\providecommand \EOS [0]{\spacefactor3000\relax}%
\providecommand \BibitemShut  [1]{\csname bibitem#1\endcsname}%
\let\auto@bib@innerbib\@empty
\bibitem [{\citenamefont {Fukuda}\ \emph {et~al.}(1998)\citenamefont {Fukuda}
  \emph {et~al.}}]{Super-Kamiokande:1998kpq}%
  \BibitemOpen
  \bibfield  {author} {\bibinfo {author} {\bibfnamefont {Y.}~\bibnamefont
  {Fukuda}} \emph {et~al.} (\bibinfo {collaboration} {Super-Kamiokande}),\
  }\href {https://doi.org/10.1103/PhysRevLett.81.1562} {\bibfield  {journal}
  {\bibinfo  {journal} {Phys. Rev. Lett.}\ }\textbf {\bibinfo {volume} {81}},\
  \bibinfo {pages} {1562} (\bibinfo {year} {1998})},\ \Eprint
  {https://arxiv.org/abs/hep-ex/9807003} {arXiv:hep-ex/9807003} \BibitemShut
  {NoStop}%
\bibitem [{\citenamefont {Ahmad}\ \emph {et~al.}(2002)\citenamefont {Ahmad}
  \emph {et~al.}}]{SNO:2002tuh}%
  \BibitemOpen
  \bibfield  {author} {\bibinfo {author} {\bibfnamefont {Q.~R.}\ \bibnamefont
  {Ahmad}} \emph {et~al.} (\bibinfo {collaboration} {SNO}),\ }\href
  {https://doi.org/10.1103/PhysRevLett.89.011301} {\bibfield  {journal}
  {\bibinfo  {journal} {Phys. Rev. Lett.}\ }\textbf {\bibinfo {volume} {89}},\
  \bibinfo {pages} {011301} (\bibinfo {year} {2002})},\ \Eprint
  {https://arxiv.org/abs/nucl-ex/0204008} {arXiv:nucl-ex/0204008} \BibitemShut
  {NoStop}%
\bibitem [{\citenamefont {Workman}\ \emph {et~al.}(2022)\citenamefont {Workman}
  \emph {et~al.}}]{ParticleDataGroup:2022pth}%
  \BibitemOpen
  \bibfield  {author} {\bibinfo {author} {\bibfnamefont {R.~L.}\ \bibnamefont
  {Workman}} \emph {et~al.} (\bibinfo {collaboration} {Particle Data Group}),\
  }\href {https://doi.org/10.1093/ptep/ptac097} {\bibfield  {journal} {\bibinfo
   {journal} {Prog. Theor. Exp. Phys.}\ }\textbf {\bibinfo {volume} {2022}},\
  \bibinfo {pages} {083C01} (\bibinfo {year} {2022})}\BibitemShut {NoStop}%
\bibitem [{\citenamefont {Petcov}(2013)}]{Petcov:2013poa}%
  \BibitemOpen
  \bibfield  {author} {\bibinfo {author} {\bibfnamefont {S.~T.}\ \bibnamefont
  {Petcov}},\ }\href {https://doi.org/10.1155/2013/852987} {\bibfield
  {journal} {\bibinfo  {journal} {Adv. High Energy Phys.}\ }\textbf {\bibinfo
  {volume} {2013}},\ \bibinfo {pages} {852987} (\bibinfo {year} {2013})},\
  \Eprint {https://arxiv.org/abs/1303.5819} {arXiv:1303.5819 [hep-ph]}
  \BibitemShut {NoStop}%
\bibitem [{\citenamefont {Dolinski}\ \emph {et~al.}(2019)\citenamefont
  {Dolinski}, \citenamefont {Poon},\ and\ \citenamefont
  {Rodejohann}}]{Dolinski:2019nrj}%
  \BibitemOpen
  \bibfield  {author} {\bibinfo {author} {\bibfnamefont {M.~J.}\ \bibnamefont
  {Dolinski}}, \bibinfo {author} {\bibfnamefont {A.~W.~P.}\ \bibnamefont
  {Poon}},\ and\ \bibinfo {author} {\bibfnamefont {W.}~\bibnamefont
  {Rodejohann}},\ }\href {https://doi.org/10.1146/annurev-nucl-101918-023407}
  {\bibfield  {journal} {\bibinfo  {journal} {Ann.\ Rev.\ Nucl.\ Part.\ Sci.}\
  }\textbf {\bibinfo {volume} {69}},\ \bibinfo {pages} {219} (\bibinfo {year}
  {2019})},\ \Eprint {https://arxiv.org/abs/1902.04097} {arXiv:1902.04097
  [nucl-ex]} \BibitemShut {NoStop}%
\bibitem [{\citenamefont {Abe}\ \emph {et~al.}(2023)\citenamefont {Abe} \emph
  {et~al.}}]{KamLAND-Zen:2022tow}%
  \BibitemOpen
  \bibfield  {author} {\bibinfo {author} {\bibfnamefont {S.}~\bibnamefont
  {Abe}} \emph {et~al.} (\bibinfo {collaboration} {KamLAND-Zen}),\ }\href
  {https://doi.org/10.1103/PhysRevLett.130.051801} {\bibfield  {journal}
  {\bibinfo  {journal} {Phys. Rev. Lett.}\ }\textbf {\bibinfo {volume} {130}},\
  \bibinfo {pages} {051801} (\bibinfo {year} {2023})},\ \Eprint
  {https://arxiv.org/abs/2203.02139} {arXiv:2203.02139 [hep-ex]} \BibitemShut
  {NoStop}%
\bibitem [{\citenamefont {Agostini}\ \emph {et~al.}(2020)\citenamefont
  {Agostini} \emph {et~al.}}]{GERDA:2020xhi}%
  \BibitemOpen
  \bibfield  {author} {\bibinfo {author} {\bibfnamefont {M.}~\bibnamefont
  {Agostini}} \emph {et~al.} (\bibinfo {collaboration} {GERDA}),\ }\href
  {https://doi.org/10.1103/PhysRevLett.125.252502} {\bibfield  {journal}
  {\bibinfo  {journal} {Phys. Rev. Lett.}\ }\textbf {\bibinfo {volume} {125}},\
  \bibinfo {pages} {252502} (\bibinfo {year} {2020})},\ \Eprint
  {https://arxiv.org/abs/2009.06079} {arXiv:2009.06079 [nucl-ex]} \BibitemShut
  {NoStop}%
\bibitem [{\citenamefont {Arnquist}\ \emph {et~al.}(2023)\citenamefont
  {Arnquist} \emph {et~al.}}]{Majorana:2022udl}%
  \BibitemOpen
  \bibfield  {author} {\bibinfo {author} {\bibfnamefont {I.~J.}\ \bibnamefont
  {Arnquist}} \emph {et~al.} (\bibinfo {collaboration} {Majorana}),\ }\href
  {https://doi.org/10.1103/PhysRevLett.130.062501} {\bibfield  {journal}
  {\bibinfo  {journal} {Phys. Rev. Lett.}\ }\textbf {\bibinfo {volume} {130}},\
  \bibinfo {pages} {062501} (\bibinfo {year} {2023})},\ \Eprint
  {https://arxiv.org/abs/2207.07638} {arXiv:2207.07638 [nucl-ex]} \BibitemShut
  {NoStop}%
\bibitem [{\citenamefont {Anton}\ \emph {et~al.}(2019)\citenamefont {Anton}
  \emph {et~al.}}]{EXO-200:2019rkq}%
  \BibitemOpen
  \bibfield  {author} {\bibinfo {author} {\bibfnamefont {G.}~\bibnamefont
  {Anton}} \emph {et~al.} (\bibinfo {collaboration} {EXO-200}),\ }\href
  {https://doi.org/10.1103/PhysRevLett.123.161802} {\bibfield  {journal}
  {\bibinfo  {journal} {Phys. Rev. Lett.}\ }\textbf {\bibinfo {volume} {123}},\
  \bibinfo {pages} {161802} (\bibinfo {year} {2019})},\ \Eprint
  {https://arxiv.org/abs/1906.02723} {arXiv:1906.02723 [hep-ex]} \BibitemShut
  {NoStop}%
\bibitem [{\citenamefont {Adams}\ \emph {et~al.}(2022)\citenamefont {Adams}
  \emph {et~al.}}]{CUORE:2021mvw}%
  \BibitemOpen
  \bibfield  {author} {\bibinfo {author} {\bibfnamefont {D.~Q.}\ \bibnamefont
  {Adams}} \emph {et~al.} (\bibinfo {collaboration} {CUORE}),\ }\href
  {https://doi.org/10.1038/s41586-022-04497-4} {\bibfield  {journal} {\bibinfo
  {journal} {Nature}\ }\textbf {\bibinfo {volume} {604}},\ \bibinfo {pages}
  {53} (\bibinfo {year} {2022})},\ \Eprint {https://arxiv.org/abs/2104.06906}
  {arXiv:2104.06906 [nucl-ex]} \BibitemShut {NoStop}%
\bibitem [{\citenamefont {Aghanim}\ \emph {et~al.}(2020)\citenamefont {Aghanim}
  \emph {et~al.}}]{Planck:2018vyg}%
  \BibitemOpen
  \bibfield  {author} {\bibinfo {author} {\bibfnamefont {N.}~\bibnamefont
  {Aghanim}} \emph {et~al.} (\bibinfo {collaboration} {Planck}),\ }\href
  {https://doi.org/10.1051/0004-6361/201833910} {\bibfield  {journal} {\bibinfo
   {journal} {Astron. Astrophys.}\ }\textbf {\bibinfo {volume} {641}},\
  \bibinfo {pages} {A6} (\bibinfo {year} {2020})},\ \bibinfo {note} {[Erratum:
  Astron.Astrophys. 652, C4 (2021)]},\ \Eprint
  {https://arxiv.org/abs/1807.06209} {arXiv:1807.06209 [astro-ph.CO]}
  \BibitemShut {NoStop}%
\bibitem [{\citenamefont {Alam}\ \emph {et~al.}(2021)\citenamefont {Alam} \emph
  {et~al.}}]{eBOSS:2020yzd}%
  \BibitemOpen
  \bibfield  {author} {\bibinfo {author} {\bibfnamefont {S.}~\bibnamefont
  {Alam}} \emph {et~al.} (\bibinfo {collaboration} {eBOSS}),\ }\href
  {https://doi.org/10.1103/PhysRevD.103.083533} {\bibfield  {journal} {\bibinfo
   {journal} {Phys. Rev. D}\ }\textbf {\bibinfo {volume} {103}},\ \bibinfo
  {pages} {083533} (\bibinfo {year} {2021})},\ \Eprint
  {https://arxiv.org/abs/2007.08991} {arXiv:2007.08991 [astro-ph.CO]}
  \BibitemShut {NoStop}%
\bibitem [{\citenamefont {Lesgourgues}\ and\ \citenamefont
  {Pastor}(2006)}]{Lesgourgues:2006nd}%
  \BibitemOpen
  \bibfield  {author} {\bibinfo {author} {\bibfnamefont {J.}~\bibnamefont
  {Lesgourgues}}\ and\ \bibinfo {author} {\bibfnamefont {S.}~\bibnamefont
  {Pastor}},\ }\href {https://doi.org/10.1016/j.physrep.2006.04.001} {\bibfield
   {journal} {\bibinfo  {journal} {Phys. Rept.}\ }\textbf {\bibinfo {volume}
  {429}},\ \bibinfo {pages} {307} (\bibinfo {year} {2006})},\ \Eprint
  {https://arxiv.org/abs/astro-ph/0603494} {arXiv:astro-ph/0603494}
  \BibitemShut {NoStop}%
\bibitem [{\citenamefont {Di~Valentino}\ \emph {et~al.}(2015)\citenamefont
  {Di~Valentino}, \citenamefont {Melchiorri},\ and\ \citenamefont
  {Silk}}]{DiValentino:2015ola}%
  \BibitemOpen
  \bibfield  {author} {\bibinfo {author} {\bibfnamefont {E.}~\bibnamefont
  {Di~Valentino}}, \bibinfo {author} {\bibfnamefont {A.}~\bibnamefont
  {Melchiorri}},\ and\ \bibinfo {author} {\bibfnamefont {J.}~\bibnamefont
  {Silk}},\ }\href {https://doi.org/10.1103/PhysRevD.92.121302} {\bibfield
  {journal} {\bibinfo  {journal} {Phys. Rev. D}\ }\textbf {\bibinfo {volume}
  {92}},\ \bibinfo {pages} {121302} (\bibinfo {year} {2015})},\ \Eprint
  {https://arxiv.org/abs/1507.06646} {arXiv:1507.06646 [astro-ph.CO]}
  \BibitemShut {NoStop}%
\bibitem [{\citenamefont {Riess}\ \emph {et~al.}(2022)\citenamefont {Riess}
  \emph {et~al.}}]{Riess:2021jrx}%
  \BibitemOpen
  \bibfield  {author} {\bibinfo {author} {\bibfnamefont {A.~G.}\ \bibnamefont
  {Riess}} \emph {et~al.},\ }\href {https://doi.org/10.3847/2041-8213/ac5c5b}
  {\bibfield  {journal} {\bibinfo  {journal} {Astrophys. J. Lett.}\ }\textbf
  {\bibinfo {volume} {934}},\ \bibinfo {pages} {L7} (\bibinfo {year} {2022})},\
  \Eprint {https://arxiv.org/abs/2112.04510} {arXiv:2112.04510 [astro-ph.CO]}
  \BibitemShut {NoStop}%
\bibitem [{\citenamefont {Knox}\ and\ \citenamefont
  {Millea}(2020)}]{Knox:2019rjx}%
  \BibitemOpen
  \bibfield  {author} {\bibinfo {author} {\bibfnamefont {L.}~\bibnamefont
  {Knox}}\ and\ \bibinfo {author} {\bibfnamefont {M.}~\bibnamefont {Millea}},\
  }\href {https://doi.org/10.1103/PhysRevD.101.043533} {\bibfield  {journal}
  {\bibinfo  {journal} {Phys. Rev. D}\ }\textbf {\bibinfo {volume} {101}},\
  \bibinfo {pages} {043533} (\bibinfo {year} {2020})},\ \Eprint
  {https://arxiv.org/abs/1908.03663} {arXiv:1908.03663 [astro-ph.CO]}
  \BibitemShut {NoStop}%
\bibitem [{\citenamefont {Abdalla}\ \emph {et~al.}(2022)\citenamefont {Abdalla}
  \emph {et~al.}}]{Abdalla:2022yfr}%
  \BibitemOpen
  \bibfield  {author} {\bibinfo {author} {\bibfnamefont {E.}~\bibnamefont
  {Abdalla}} \emph {et~al.},\ }\href
  {https://doi.org/10.1016/j.jheap.2022.04.002} {\bibfield  {journal} {\bibinfo
   {journal} {JHEAp}\ }\textbf {\bibinfo {volume} {34}},\ \bibinfo {pages} {49}
  (\bibinfo {year} {2022})},\ \Eprint {https://arxiv.org/abs/2203.06142}
  {arXiv:2203.06142 [astro-ph.CO]} \BibitemShut {NoStop}%
\bibitem [{\citenamefont {Drexlin}\ \emph {et~al.}(2013)\citenamefont
  {Drexlin}, \citenamefont {Hannen}, \citenamefont {Mertens},\ and\
  \citenamefont {Weinheimer}}]{Drexlin:2013lha}%
  \BibitemOpen
  \bibfield  {author} {\bibinfo {author} {\bibfnamefont {G.}~\bibnamefont
  {Drexlin}}, \bibinfo {author} {\bibfnamefont {V.}~\bibnamefont {Hannen}},
  \bibinfo {author} {\bibfnamefont {S.}~\bibnamefont {Mertens}},\ and\ \bibinfo
  {author} {\bibfnamefont {C.}~\bibnamefont {Weinheimer}},\ }\href
  {https://doi.org/10.1155/2013/293986} {\bibfield  {journal} {\bibinfo
  {journal} {Adv. High Energy Phys.}\ }\textbf {\bibinfo {volume} {2013}},\
  \bibinfo {pages} {293986} (\bibinfo {year} {2013})},\ \Eprint
  {https://arxiv.org/abs/1307.0101} {arXiv:1307.0101 [physics.ins-det]}
  \BibitemShut {NoStop}%
\bibitem [{\citenamefont {Formaggio}\ \emph {et~al.}(2021)\citenamefont
  {Formaggio}, \citenamefont {de~Gouv\^ea},\ and\ \citenamefont
  {Robertson}}]{Formaggio:2021nfz}%
  \BibitemOpen
  \bibfield  {author} {\bibinfo {author} {\bibfnamefont {J.~A.}\ \bibnamefont
  {Formaggio}}, \bibinfo {author} {\bibfnamefont {A.~L.~C.}\ \bibnamefont
  {de~Gouv\^ea}},\ and\ \bibinfo {author} {\bibfnamefont {R.~G.~H.}\
  \bibnamefont {Robertson}},\ }\href
  {https://doi.org/10.1016/j.physrep.2021.02.002} {\bibfield  {journal}
  {\bibinfo  {journal} {Phys. Rept.}\ }\textbf {\bibinfo {volume} {914}},\
  \bibinfo {pages} {1} (\bibinfo {year} {2021})},\ \Eprint
  {https://arxiv.org/abs/2102.00594} {arXiv:2102.00594 [nucl-ex]} \BibitemShut
  {NoStop}%
\bibitem [{\citenamefont {Pontecorvo}(1957)}]{Pontecorvo:1957cp}%
  \BibitemOpen
  \bibfield  {author} {\bibinfo {author} {\bibfnamefont {B.}~\bibnamefont
  {Pontecorvo}},\ }\href@noop {} {\bibfield  {journal} {\bibinfo  {journal}
  {Sov. Phys. JETP}\ }\textbf {\bibinfo {volume} {6}},\ \bibinfo {pages} {429}
  (\bibinfo {year} {1957})}\BibitemShut {NoStop}%
\bibitem [{\citenamefont {Maki}\ \emph {et~al.}(1962)\citenamefont {Maki},
  \citenamefont {Nakagawa},\ and\ \citenamefont {Sakata}}]{Maki:1962mu}%
  \BibitemOpen
  \bibfield  {author} {\bibinfo {author} {\bibfnamefont {Z.}~\bibnamefont
  {Maki}}, \bibinfo {author} {\bibfnamefont {M.}~\bibnamefont {Nakagawa}},\
  and\ \bibinfo {author} {\bibfnamefont {S.}~\bibnamefont {Sakata}},\ }\href
  {https://doi.org/10.1143/PTP.28.870} {\bibfield  {journal} {\bibinfo
  {journal} {Prog. Theor. Phys.}\ }\textbf {\bibinfo {volume} {28}},\ \bibinfo
  {pages} {870} (\bibinfo {year} {1962})}\BibitemShut {NoStop}%
\bibitem [{\citenamefont {Aker}\ \emph {et~al.}(2022)\citenamefont {Aker} \emph
  {et~al.}}]{KATRIN:2021uub}%
  \BibitemOpen
  \bibfield  {author} {\bibinfo {author} {\bibfnamefont {M.}~\bibnamefont
  {Aker}} \emph {et~al.} (\bibinfo {collaboration} {KATRIN}),\ }\href
  {https://doi.org/10.1038/s41567-021-01463-1} {\bibfield  {journal} {\bibinfo
  {journal} {Nature Phys.}\ }\textbf {\bibinfo {volume} {18}},\ \bibinfo
  {pages} {160} (\bibinfo {year} {2022})},\ \Eprint
  {https://arxiv.org/abs/2105.08533} {arXiv:2105.08533 [hep-ex]} \BibitemShut
  {NoStop}%
\bibitem [{\citenamefont {Saenz}\ \emph {et~al.}(2000)\citenamefont {Saenz},
  \citenamefont {Jonsell},\ and\ \citenamefont {Froelich}}]{Saenz:2000dul}%
  \BibitemOpen
  \bibfield  {author} {\bibinfo {author} {\bibfnamefont {A.}~\bibnamefont
  {Saenz}}, \bibinfo {author} {\bibfnamefont {S.}~\bibnamefont {Jonsell}},\
  and\ \bibinfo {author} {\bibfnamefont {P.}~\bibnamefont {Froelich}},\ }\href
  {https://doi.org/10.1103/PhysRevLett.84.242} {\bibfield  {journal} {\bibinfo
  {journal} {Phys. Rev. Lett.}\ }\textbf {\bibinfo {volume} {84}},\ \bibinfo
  {pages} {242} (\bibinfo {year} {2000})}\BibitemShut {NoStop}%
\bibitem [{\citenamefont {Bodine}\ \emph {et~al.}(2015)\citenamefont {Bodine},
  \citenamefont {Parno},\ and\ \citenamefont {Robertson}}]{Bodine:2015sma}%
  \BibitemOpen
  \bibfield  {author} {\bibinfo {author} {\bibfnamefont {L.~I.}\ \bibnamefont
  {Bodine}}, \bibinfo {author} {\bibfnamefont {D.~S.}\ \bibnamefont {Parno}},\
  and\ \bibinfo {author} {\bibfnamefont {R.~G.~H.}\ \bibnamefont {Robertson}},\
  }\href {https://doi.org/10.1103/PhysRevC.91.035505} {\bibfield  {journal}
  {\bibinfo  {journal} {Phys. Rev. C}\ }\textbf {\bibinfo {volume} {91}},\
  \bibinfo {pages} {035505} (\bibinfo {year} {2015})},\ \Eprint
  {https://arxiv.org/abs/1502.03497} {arXiv:1502.03497 [nucl-ex]} \BibitemShut
  {NoStop}%
\bibitem [{\citenamefont {Gastaldo}\ \emph {et~al.}(2017)\citenamefont
  {Gastaldo} \emph {et~al.}}]{Gastaldo:2017edk}%
  \BibitemOpen
  \bibfield  {author} {\bibinfo {author} {\bibfnamefont {L.}~\bibnamefont
  {Gastaldo}} \emph {et~al.},\ }\href
  {https://doi.org/10.1140/epjst/e2017-70071-y} {\bibfield  {journal} {\bibinfo
   {journal} {Eur. Phys. J. ST}\ }\textbf {\bibinfo {volume} {226}},\ \bibinfo
  {pages} {1623} (\bibinfo {year} {2017})}\BibitemShut {NoStop}%
\bibitem [{\citenamefont {Alpert}\ \emph {et~al.}(2015)\citenamefont {Alpert}
  \emph {et~al.}}]{Alpert:2014lfa}%
  \BibitemOpen
  \bibfield  {author} {\bibinfo {author} {\bibfnamefont {B.}~\bibnamefont
  {Alpert}} \emph {et~al.},\ }\href
  {https://doi.org/10.1140/epjc/s10052-015-3329-5} {\bibfield  {journal}
  {\bibinfo  {journal} {Eur. Phys. J. C}\ }\textbf {\bibinfo {volume} {75}},\
  \bibinfo {pages} {112} (\bibinfo {year} {2015})},\ \Eprint
  {https://arxiv.org/abs/1412.5060} {arXiv:1412.5060 [physics.ins-det]}
  \BibitemShut {NoStop}%
\bibitem [{\citenamefont {Velte}\ \emph {et~al.}(2019)\citenamefont {Velte}
  \emph {et~al.}}]{Velte:2019jvx}%
  \BibitemOpen
  \bibfield  {author} {\bibinfo {author} {\bibfnamefont {C.}~\bibnamefont
  {Velte}} \emph {et~al.},\ }\href
  {https://doi.org/10.1140/epjc/s10052-019-7513-x} {\bibfield  {journal}
  {\bibinfo  {journal} {Eur. Phys. J. C}\ }\textbf {\bibinfo {volume} {79}},\
  \bibinfo {pages} {1026} (\bibinfo {year} {2019})}\BibitemShut {NoStop}%
\bibitem [{\citenamefont {Monreal}\ and\ \citenamefont
  {Formaggio}(2009)}]{Monreal:2009za}%
  \BibitemOpen
  \bibfield  {author} {\bibinfo {author} {\bibfnamefont {B.}~\bibnamefont
  {Monreal}}\ and\ \bibinfo {author} {\bibfnamefont {J.~A.}\ \bibnamefont
  {Formaggio}},\ }\href {https://doi.org/10.1103/PhysRevD.80.051301} {\bibfield
   {journal} {\bibinfo  {journal} {Phys.\ Rev.\ D}\ }\textbf {\bibinfo {volume}
  {80}},\ \bibinfo {pages} {051301} (\bibinfo {year} {2009})},\ \Eprint
  {https://arxiv.org/abs/0904.2860} {arXiv:0904.2860 [nucl-ex]} \BibitemShut
  {NoStop}%
\bibitem [{\citenamefont {Asner}\ \emph {et~al.}(2015)\citenamefont {Asner}
  \emph {et~al.}}]{Asner:2014cwa}%
  \BibitemOpen
  \bibfield  {author} {\bibinfo {author} {\bibfnamefont {D.~M.}\ \bibnamefont
  {Asner}} \emph {et~al.} (\bibinfo {collaboration} {Project 8}),\ }\href
  {https://doi.org/10.1103/PhysRevLett.114.162501} {\bibfield  {journal}
  {\bibinfo  {journal} {Phys.\ Rev.\ Lett.}\ }\textbf {\bibinfo {volume}
  {114}},\ \bibinfo {pages} {162501} (\bibinfo {year} {2015})},\ \Eprint
  {https://arxiv.org/abs/1408.5362} {arXiv:1408.5362 [physics.ins-det]}
  \BibitemShut {NoStop}%
\bibitem [{\citenamefont {Byron}\ \emph {et~al.}(2022)\citenamefont {Byron}
  \emph {et~al.}}]{Byron:2022wtr}%
  \BibitemOpen
  \bibfield  {author} {\bibinfo {author} {\bibfnamefont {W.}~\bibnamefont
  {Byron}} \emph {et~al.},\ }\Eprint {https://arxiv.org/abs/2209.02870}
  {arXiv:2209.02870 [nucl-ex]}  (\bibinfo {year} {2022})\BibitemShut {NoStop}%
\bibitem [{\citenamefont {Ashtari~Esfahani}\ \emph {et~al.}(2017)\citenamefont
  {Ashtari~Esfahani} \emph {et~al.}}]{Esfahani:2017dmu}%
  \BibitemOpen
  \bibfield  {author} {\bibinfo {author} {\bibfnamefont {A.}~\bibnamefont
  {Ashtari~Esfahani}} \emph {et~al.} (\bibinfo {collaboration} {Project 8}),\
  }\href {https://doi.org/10.1088/1361-6471/aa5b4f} {\bibfield  {journal}
  {\bibinfo  {journal} {J.\ Phys.\ G}\ }\textbf {\bibinfo {volume} {44}},\
  \bibinfo {pages} {054004} (\bibinfo {year} {2017})},\ \Eprint
  {https://arxiv.org/abs/1703.02037} {arXiv:1703.02037 [physics.ins-det]}
  \BibitemShut {NoStop}%
\bibitem [{\citenamefont {Aker}\ \emph {et~al.}(2021)\citenamefont {Aker} \emph
  {et~al.}}]{KATRIN:2021fgc}%
  \BibitemOpen
  \bibfield  {author} {\bibinfo {author} {\bibfnamefont {M.}~\bibnamefont
  {Aker}} \emph {et~al.} (\bibinfo {collaboration} {KATRIN}),\ }\href
  {https://doi.org/10.1103/PhysRevD.104.012005} {\bibfield  {journal} {\bibinfo
   {journal} {Phys. Rev. D}\ }\textbf {\bibinfo {volume} {104}},\ \bibinfo
  {pages} {012005} (\bibinfo {year} {2021})},\ \Eprint
  {https://arxiv.org/abs/2101.05253} {arXiv:2101.05253 [hep-ex]} \BibitemShut
  {NoStop}%
\bibitem [{\citenamefont {Esfahani}\ \emph {et~al.}(2022)\citenamefont
  {Esfahani} \emph {et~al.}}]{Project8:2022wqh}%
  \BibitemOpen
  \bibfield  {author} {\bibinfo {author} {\bibfnamefont {A.~A.}\ \bibnamefont
  {Esfahani}} \emph {et~al.} (\bibinfo {collaboration} {Project 8}),\ }in\
  \href@noop {} {\emph {\bibinfo {booktitle} {{2022 Snowmass Summer Study}}}}\
  (\bibinfo {year} {2022})\ \Eprint {https://arxiv.org/abs/2203.07349}
  {arXiv:2203.07349 [nucl-ex]} \BibitemShut {NoStop}%
\bibitem [{\citenamefont {Ashtari~Esfahani}\ \emph {et~al.}(2022)\citenamefont
  {Ashtari~Esfahani} \emph {et~al.}}]{tritiumPRC:2022}%
  \BibitemOpen
  \bibfield  {author} {\bibinfo {author} {\bibfnamefont {A.}~\bibnamefont
  {Ashtari~Esfahani}} \emph {et~al.} (\bibinfo {collaboration} {Project 8})}
  (\bibinfo {year} {2022}),\ \bibinfo {note} {long-form article in
  preparation}\BibitemShut {NoStop}%
\bibitem [{\citenamefont {V{\'e}nos}\ \emph {et~al.}(2005)\citenamefont
  {V{\'e}nos}, \citenamefont {Spalek}, \citenamefont {Lebeda},\ and\
  \citenamefont {Fiser}}]{Venos:2005vn}%
  \BibitemOpen
  \bibfield  {author} {\bibinfo {author} {\bibfnamefont {D.}~\bibnamefont
  {V{\'e}nos}}, \bibinfo {author} {\bibfnamefont {A.}~\bibnamefont {Spalek}},
  \bibinfo {author} {\bibfnamefont {O.}~\bibnamefont {Lebeda}},\ and\ \bibinfo
  {author} {\bibfnamefont {M.}~\bibnamefont {Fiser}},\ }\href@noop {}
  {\bibfield  {journal} {\bibinfo  {journal} {Applied Radiation and Isotopes}\
  }\textbf {\bibinfo {volume} {63}},\ \bibinfo {pages} {323} (\bibinfo {year}
  {2005})}\BibitemShut {NoStop}%
\bibitem [{\citenamefont {{Hickish}}\ \emph {et~al.}(2016)\citenamefont
  {{Hickish}} \emph {et~al.}}]{Hickish2016}%
  \BibitemOpen
  \bibfield  {author} {\bibinfo {author} {\bibfnamefont {J.}~\bibnamefont
  {{Hickish}}} \emph {et~al.},\ }\href
  {https://doi.org/10.1142/S2251171716410014} {\bibfield  {journal} {\bibinfo
  {journal} {Journal of Astronomical Instrumentation}\ }\textbf {\bibinfo
  {volume} {5}},\ \bibinfo {eid} {1641001-12} (\bibinfo {year} {2016})},\
  \Eprint {https://arxiv.org/abs/1611.01826} {arXiv:1611.01826 [astro-ph.IM]}
  \BibitemShut {NoStop}%
\bibitem [{\citenamefont {Ashtari~Esfahani}\ \emph
  {et~al.}(2019{\natexlab{a}})\citenamefont {Ashtari~Esfahani} \emph
  {et~al.}}]{Esfahani:2019mpr}%
  \BibitemOpen
  \bibfield  {author} {\bibinfo {author} {\bibfnamefont {A.}~\bibnamefont
  {Ashtari~Esfahani}} \emph {et~al.},\ }\href
  {https://doi.org/10.1103/PhysRevC.99.055501} {\bibfield  {journal} {\bibinfo
  {journal} {Phys.\ Rev.\ C}\ }\textbf {\bibinfo {volume} {99}},\ \bibinfo
  {pages} {055501} (\bibinfo {year} {2019}{\natexlab{a}})},\ \Eprint
  {https://arxiv.org/abs/1901.02844} {arXiv:1901.02844 [physics.ins-det]}
  \BibitemShut {NoStop}%
\bibitem [{\citenamefont {Ashtari~Esfahani}\ \emph
  {et~al.}(2019{\natexlab{b}})\citenamefont {Ashtari~Esfahani} \emph
  {et~al.}}]{AshtariEsfahani:2019mwv}%
  \BibitemOpen
  \bibfield  {author} {\bibinfo {author} {\bibfnamefont {A.}~\bibnamefont
  {Ashtari~Esfahani}} \emph {et~al.} (\bibinfo {collaboration} {Project 8}),\
  }\href {https://doi.org/10.1088/1367-2630/ab550d} {\bibfield  {journal}
  {\bibinfo  {journal} {New J.\ Phys.}\ }\textbf {\bibinfo {volume} {21}},\
  \bibinfo {pages} {113051} (\bibinfo {year} {2019}{\natexlab{b}})},\ \Eprint
  {https://arxiv.org/abs/1907.11124} {arXiv:1907.11124 [physics.comp-ph]}
  \BibitemShut {NoStop}%
\bibitem [{\citenamefont {Robertson}\ and\ \citenamefont
  {Venkatapathy}(2020)}]{Robertson:2020dd}%
  \BibitemOpen
  \bibfield  {author} {\bibinfo {author} {\bibfnamefont {R.~G.~H.}\
  \bibnamefont {Robertson}}\ and\ \bibinfo {author} {\bibfnamefont
  {V.}~\bibnamefont {Venkatapathy}},\ }\href@noop {} {\bibfield  {journal}
  {\bibinfo  {journal} {Phys. Rev.}\ }\textbf {\bibinfo {volume} {102}},\
  \bibinfo {pages} {035502} (\bibinfo {year} {2020})}\BibitemShut {NoStop}%
\bibitem [{\citenamefont {Baker}\ and\ \citenamefont
  {Cousins}(1984)}]{Baker:1983tu}%
  \BibitemOpen
  \bibfield  {author} {\bibinfo {author} {\bibfnamefont {S.}~\bibnamefont
  {Baker}}\ and\ \bibinfo {author} {\bibfnamefont {R.~D.}\ \bibnamefont
  {Cousins}},\ }\href {https://doi.org/10.1016/0167-5087(84)90016-4} {\bibfield
   {journal} {\bibinfo  {journal} {Nucl. Instrum. Meth.}\ }\textbf {\bibinfo
  {volume} {221}},\ \bibinfo {pages} {437} (\bibinfo {year}
  {1984})}\BibitemShut {NoStop}%
\bibitem [{\citenamefont {Altenm\"uller}\ \emph {et~al.}(2020)\citenamefont
  {Altenm\"uller} \emph {et~al.}}]{Altenmuller:2019ddl}%
  \BibitemOpen
  \bibfield  {author} {\bibinfo {author} {\bibfnamefont {K.}~\bibnamefont
  {Altenm\"uller}} \emph {et~al.},\ }\href
  {https://doi.org/10.1088/1361-6471/ab8480} {\bibfield  {journal} {\bibinfo
  {journal} {J. Phys. G}\ }\textbf {\bibinfo {volume} {47}},\ \bibinfo {pages}
  {065002} (\bibinfo {year} {2020})},\ \Eprint
  {https://arxiv.org/abs/1903.06452} {arXiv:1903.06452 [physics.ins-det]}
  \BibitemShut {NoStop}%
\bibitem [{\citenamefont {Purcell}(1946)}]{Purcell1946}%
  \BibitemOpen
  \bibfield  {author} {\bibinfo {author} {\bibfnamefont {E.~M.}\ \bibnamefont
  {Purcell}},\ }\href {https://doi.org/10.1103/PhysRev.69.674.2} {\bibfield
  {journal} {\bibinfo  {journal} {Phys. Rev.}\ }\textbf {\bibinfo {volume}
  {69}},\ \bibinfo {pages} {681} (\bibinfo {year} {1946})}\BibitemShut
  {NoStop}%
\bibitem [{\citenamefont {Gabrielse}\ and\ \citenamefont
  {Dehmelt}(1985)}]{Gabrielse:1985zz}%
  \BibitemOpen
  \bibfield  {author} {\bibinfo {author} {\bibfnamefont {G.}~\bibnamefont
  {Gabrielse}}\ and\ \bibinfo {author} {\bibfnamefont {H.}~\bibnamefont
  {Dehmelt}},\ }\href {https://doi.org/10.1103/PhysRevLett.55.67} {\bibfield
  {journal} {\bibinfo  {journal} {Phys. Rev. Lett.}\ }\textbf {\bibinfo
  {volume} {55}},\ \bibinfo {pages} {67} (\bibinfo {year} {1985})}\BibitemShut
  {NoStop}%
\bibitem [{\citenamefont {Myers}\ \emph {et~al.}(2015)\citenamefont {Myers},
  \citenamefont {Wagner}, \citenamefont {Kracke},\ and\ \citenamefont
  {Wesson}}]{Myers:2015lca}%
  \BibitemOpen
  \bibfield  {author} {\bibinfo {author} {\bibfnamefont {E.~G.}\ \bibnamefont
  {Myers}}, \bibinfo {author} {\bibfnamefont {A.}~\bibnamefont {Wagner}},
  \bibinfo {author} {\bibfnamefont {H.}~\bibnamefont {Kracke}},\ and\ \bibinfo
  {author} {\bibfnamefont {B.~A.}\ \bibnamefont {Wesson}},\ }\href
  {https://doi.org/10.1103/PhysRevLett.114.013003} {\bibfield  {journal}
  {\bibinfo  {journal} {Phys. Rev. Lett.}\ }\textbf {\bibinfo {volume} {114}},\
  \bibinfo {pages} {013003} (\bibinfo {year} {2015})}\BibitemShut {NoStop}%
\bibitem [{\citenamefont {Kleesiek}\ \emph {et~al.}(2019)\citenamefont
  {Kleesiek} \emph {et~al.}}]{Kleesiek:2018mel}%
  \BibitemOpen
  \bibfield  {author} {\bibinfo {author} {\bibfnamefont {M.}~\bibnamefont
  {Kleesiek}} \emph {et~al.},\ }\href
  {https://doi.org/10.1140/epjc/s10052-019-6686-7} {\bibfield  {journal}
  {\bibinfo  {journal} {Eur. Phys. J. C}\ }\textbf {\bibinfo {volume} {79}},\
  \bibinfo {pages} {204} (\bibinfo {year} {2019})},\ \Eprint
  {https://arxiv.org/abs/1806.00369} {arXiv:1806.00369 [physics.data-an]}
  \BibitemShut {NoStop}%
\bibitem [{\citenamefont {Ashtari~Esfahani}\ \emph {et~al.}(2021)\citenamefont
  {Ashtari~Esfahani} \emph {et~al.}}]{AshtariEsfahani:2021moh}%
  \BibitemOpen
  \bibfield  {author} {\bibinfo {author} {\bibfnamefont {A.}~\bibnamefont
  {Ashtari~Esfahani}} \emph {et~al.},\ }\href
  {https://doi.org/10.1103/PhysRevC.103.065501} {\bibfield  {journal} {\bibinfo
   {journal} {Phys. Rev. C}\ }\textbf {\bibinfo {volume} {103}},\ \bibinfo
  {pages} {065501} (\bibinfo {year} {2021})},\ \Eprint
  {https://arxiv.org/abs/2012.14341} {arXiv:2012.14341 [physics.data-an]}
  \BibitemShut {NoStop}%
\bibitem [{\citenamefont {Kraus}\ \emph {et~al.}(2005)\citenamefont {Kraus}
  \emph {et~al.}}]{Kraus:2004zw}%
  \BibitemOpen
  \bibfield  {author} {\bibinfo {author} {\bibfnamefont {C.}~\bibnamefont
  {Kraus}} \emph {et~al.},\ }\href {https://doi.org/10.1140/epjc/s2005-02139-7}
  {\bibfield  {journal} {\bibinfo  {journal} {Eur. Phys. J. C}\ }\textbf
  {\bibinfo {volume} {40}},\ \bibinfo {pages} {447} (\bibinfo {year} {2005})},\
  \Eprint {https://arxiv.org/abs/hep-ex/0412056} {arXiv:hep-ex/0412056}
  \BibitemShut {NoStop}%
\bibitem [{\citenamefont {Feldman}\ and\ \citenamefont
  {Cousins}(1998)}]{Feldman:1997qc}%
  \BibitemOpen
  \bibfield  {author} {\bibinfo {author} {\bibfnamefont {G.~J.}\ \bibnamefont
  {Feldman}}\ and\ \bibinfo {author} {\bibfnamefont {R.~D.}\ \bibnamefont
  {Cousins}},\ }\href {https://doi.org/10.1103/PhysRevD.57.3873} {\bibfield
  {journal} {\bibinfo  {journal} {Phys. Rev. D}\ }\textbf {\bibinfo {volume}
  {57}},\ \bibinfo {pages} {3873} (\bibinfo {year} {1998})},\ \Eprint
  {https://arxiv.org/abs/physics/9711021} {arXiv:physics/9711021} \BibitemShut
  {NoStop}%
\bibitem [{\citenamefont {Cowan}\ \emph {et~al.}(2011)\citenamefont {Cowan},
  \citenamefont {Cranmer}, \citenamefont {Gross},\ and\ \citenamefont
  {Vitells}}]{Cowan:2010js}%
  \BibitemOpen
  \bibfield  {author} {\bibinfo {author} {\bibfnamefont {G.}~\bibnamefont
  {Cowan}}, \bibinfo {author} {\bibfnamefont {K.}~\bibnamefont {Cranmer}},
  \bibinfo {author} {\bibfnamefont {E.}~\bibnamefont {Gross}},\ and\ \bibinfo
  {author} {\bibfnamefont {O.}~\bibnamefont {Vitells}},\ }\href
  {https://doi.org/10.1140/epjc/s10052-011-1554-0} {\bibfield  {journal}
  {\bibinfo  {journal} {Eur. Phys. J. C}\ }\textbf {\bibinfo {volume} {71}},\
  \bibinfo {pages} {1554} (\bibinfo {year} {2011})},\ \bibinfo {note}
  {[Erratum: Eur.Phys.J.C 73, 2501 (2013)]},\ \Eprint
  {https://arxiv.org/abs/1007.1727} {arXiv:1007.1727 [physics.data-an]}
  \BibitemShut {NoStop}%
\bibitem [{\citenamefont {Ashtari~Esfahani}\ \emph {et~al.}(2020)\citenamefont
  {Ashtari~Esfahani} \emph {et~al.}}]{AshtariEsfahani:2019mxy}%
  \BibitemOpen
  \bibfield  {author} {\bibinfo {author} {\bibfnamefont {A.}~\bibnamefont
  {Ashtari~Esfahani}} \emph {et~al.},\ }\href
  {https://doi.org/10.1088/1367-2630/ab71bd} {\bibfield  {journal} {\bibinfo
  {journal} {New J.\ Phys.}\ }\textbf {\bibinfo {volume} {22}},\ \bibinfo
  {pages} {033004} (\bibinfo {year} {2020})},\ \Eprint
  {https://arxiv.org/abs/1909.08115} {arXiv:1909.08115 [nucl-ex]} \BibitemShut
  {NoStop}%
\end{thebibliography}%

\end{document}